\documentclass[10pt,aps,pra,twocolumn, nofootinbib]{revtex4-1}
\usepackage{amssymb,amsmath}
\usepackage{natbib}
\usepackage{graphicx}
\usepackage{color}
\usepackage{hyperref}
\usepackage{listings}

\hypersetup{
    unicode=false,          
    pdftoolbar=true,        
    pdfmenubar=true,        
    pdffitwindow=false,     
    pdfstartview={FitH},    
    pdfauthor={Sean Hodgman},     
    colorlinks=true,       
    linkcolor=blue,          
    citecolor=red,        
}
\graphicspath{{figures/}}

\begin{document}

\definecolor{dkgreen}{rgb}{0,0.6,0}
\definecolor{gray}{rgb}{0.5,0.5,0.5}
\definecolor{mauve}{rgb}{0.58,0,0.82}

\lstset{frame=tb,
  	language=Matlab,
  	aboveskip=3mm,
  	belowskip=3mm,
  	showstringspaces=false,
  	columns=flexible,
  	basicstyle={\small\ttfamily},
  	numbers=none,
  	numberstyle=\tiny\color{gray},
 	keywordstyle=\color{blue},
	commentstyle=\color{dkgreen},
  	stringstyle=\color{mauve},
  	breaklines=true,
  	breakatwhitespace=true
  	tabsize=3
}

\title{Rapid generation of metastable helium Bose-Einstein condensates}
\author{A.~H.~Abbas$^{1,2}$}
\author{X.~Meng$^{1}$}
\author{R.~S.~Patil$^{1,3}$}
\altaffiliation[Current address: ]{Department of Physics, The Pennsylvania State University, University Park, Pennsylvania 16802, USA}
\author{J.~A.~Ross$^{1}$}
\author{A.~G.~Truscott$^{1}$}
\author{S.~S.~Hodgman$^1$}
\email{sean.hodgman@anu.edu.au}

\affiliation{\normalsize{$^1$ Research School of Physics, Australian National University, Canberra 0200, Australia}}
\affiliation{\normalsize{$^2$ Department of Physics, Faculty of Science, Cairo University, Giza, Egypt}}
\affiliation{\normalsize{$^3$ Indian Institute of Science Education and Research, Pune 411008, India}}

\date{\today}

\begin{abstract}
We report the realisation of Bose-Einstein condensation (BEC) of metastable helium atoms using an in-vacuum coil magnetic trap and a crossed beam optical dipole trap. A novel quadrupole-Ioffe configuration (QUIC) magnetic trap made from in-vacuum hollow copper tubes provides fast switching times while generating traps with a 10G bias, without compromising optical access.  The bias enables in-trap 1D doppler cooling to be used, which is the only cooling stage between the magneto-optic trap (MOT) and the optical dipole trap.  This allows direct transfer to the dipole trap without the need for any additional evaporative cooling in the magnetic trap. The entire experimental sequence takes 3.3 seconds, with essentially pure BECs observed with $\sim 10^{6}$ atoms after evaporative cooling in the dipole trap. 
\end{abstract}

\maketitle

\par The experimental creation of Bose-Einstein condensates (BECs) of dilute weakly interacting gases of atoms \cite{Anderson95,Bradley95,Davis95} has opened the possibility of exploring interesting phenomena of the quantum world on a macroscopic scale. Over the subsequent years the field has exploded, with BECs now used in diverse fields of quantum science, including quantum many-body systems \cite{Bloch08}, 
topological physics \cite{Cooper19} and precision inertial measurements \cite{Geiger20}.  While BEC experiments were initially limited to observing collective properties of the ensemble, a number of more recent detection techniques allow for the detection of individual atoms \cite{Ott16}, opening up a broad new range of experimental possibilities.  In experiments involving alkali or rare earth atoms, such single atom detection is usually performed via high resolution fluorescence imaging, either in-situ in optical lattices \cite{Bakr2009,Sherson2010} or after expansion from a trap \cite{Buecker09,Bergschneider18}.  However, such techniques usually have limitations on their spatial extent and the atom number able to be imaged.

An alternative method of single atom detection exploits the high internal energy of helium atoms trapped in the first atomic excited $2^3S_1$ metastable state (He$^*$), which has 19.8eV of internal energy.  The high internal energy allows direct detection of individual atoms with full 3D resolution using electronic detectors such as multi-channel plate and delay-line detectors (MCP-DLD) \cite{Vassen2012}.  This has opened up a wide range of exciting experimental possibilities, allowing quantum-optics equivalent demonstrations with atoms of iconic experiments such as the Hanbury Brown-Twiss (HBT) effect \cite{Schellekens2005}, Wheeler's delayed choice \cite{Manning2015_Wheeler}, the Hong-Ou-Mandel effect \cite{Lopes15}, ghost imaging \cite{Khakimov2016} and the measurement of Bell correlations \cite{Shin2018}.  More than just replicating quantum optics though, the atomic interactions allow effects to be seen that are not possible with photons, such as Fermionic anti-bunching \cite{Jeltes2007}, quantum depletion \cite{Chang16} and strongly interacting lattice physics \cite{Cayla20}.  

Crucial to many of these experiments are the measurement of HBT-style correlation functions - the key observable enabled by single atom detection.  However, such correlation functions require large amounts of data, especially if higher order correlations are being measured \cite{Dall2013_idealnbody,Hodgman2017}, often needing 10,000-100,000 individual experimental runs for a single experiment.  This has led to constant improvements across generations of experimental He$^*$ apparatuses aiming for ever shorter experimental sequences.  Most He$^*$ experiments \cite{Robert01,DosSantos01,Tychkov06,Dall2007,Keller14} use magnetic traps, where the sequence length is limited by the slow thermalisation rates in magnetic traps that have relatively weak trap frequencies.  While using liquid helium in the source stage can substantially reduce the sequence length \cite{Dall2007}, it is less practical for everyday operation.  The tight traps provided by dipole traps can overcome this, either in a hybrid combination with a magnetic trap \cite{Flores15} or in a stand-alone cross beam configuration \cite{Bouton15}.  However, due to the limited depth of the dipole trap, to enable an efficient transfer the atoms need to first be cooled, either optically via complex additional cooling schemes such as grey molasses and/or evaporatively in a magnetic trap \cite{Bouton15}.  

In this work, we report on the construction of a He$^*$ BEC machine capable of creating condensates in a simplified, rapid sequence that takes only 3.3s in total. An initial stage of 1D Doppler cooling \cite{Tychkov06} is performed in a biased magnetic trap, constructed using a simplified in-vacuum water-cooled coil geometry to produce a biased quadrupole trap.  This lowers the temperature enough to permit direct loading of a crossed dipole trap, which allows fast and efficient evaporative cooling.  The final BECs are produced with $10^6$ atoms.  As well as the relatively small size and minimal number of coil turns, which allow fast switch off times, a major advantage of this magnetic trap design is that the small size allows good optical access to the atomic cloud.  This will make the apparatus ideal for future optical lattice experiments \cite{Cayla20}.

\begin{figure*}[hbt!]
    \includegraphics[width=0.9\textwidth]{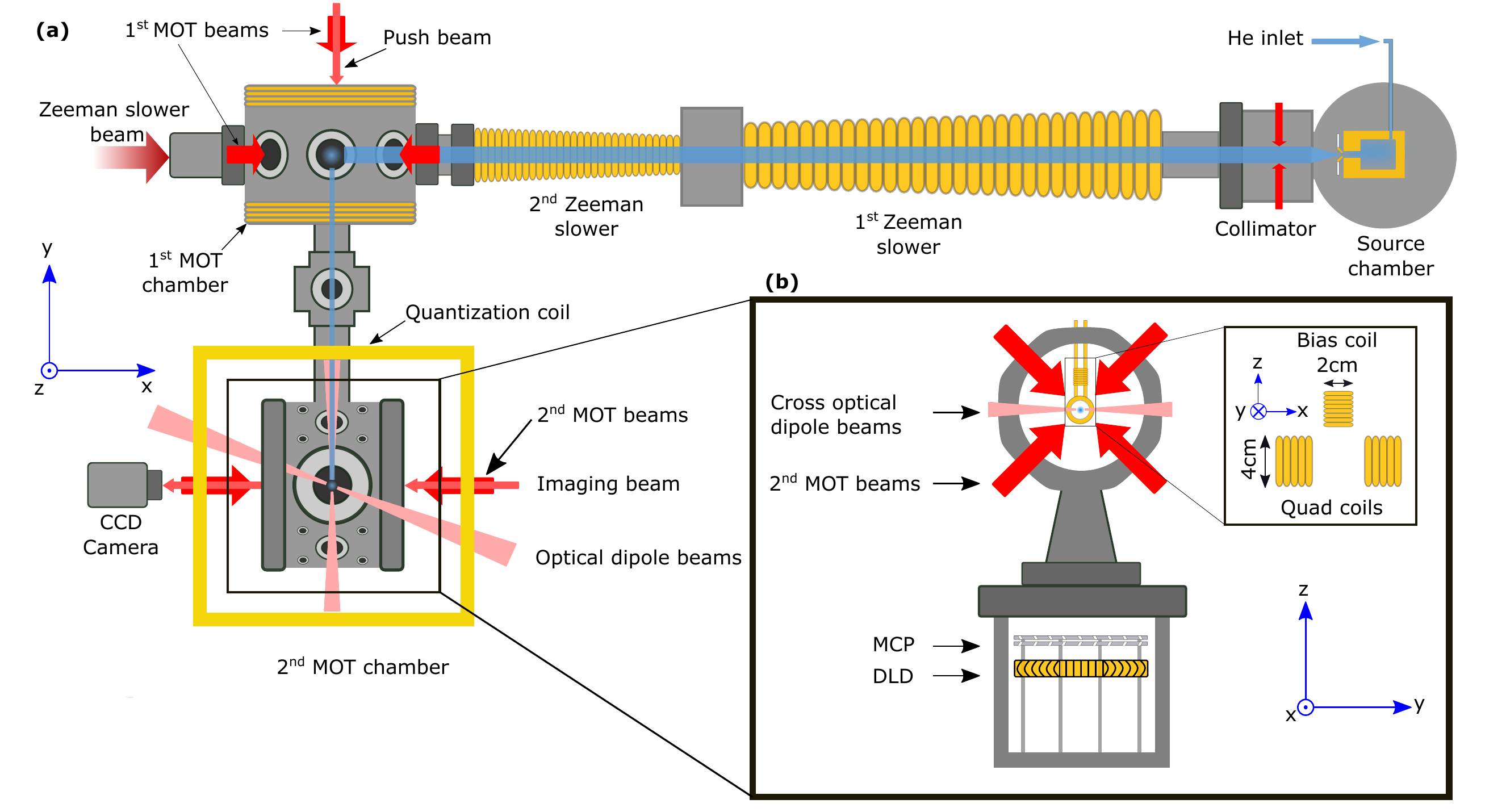}
    \caption{Schematic of the experimental apparatus (see main text for details).  (a) Top view of the experimental setup.  A beam of He$^*$ atoms is produced in a cryogenically liquid nitrogen cooled source, before being optically collimated, slowed and cooled in a 1st MOT stage. This forms a source for the $2^{nd}$ MOT, where the atoms are then transferred to a magnetic trap and subsequently to an optical dipole trap, where the BEC is generated. (b) Side view of the $2^{nd}$ MOT chamber, showing the quadruple-Ioffe configuration in-vacuum trap and MCP-DLD detector.  The inset shows the trapping coil geometry from a different angle.}
    \label{fig:1}
\end{figure*}

\begin{figure}[bt!]
    \centering
    \includegraphics[width=0.5\textwidth]{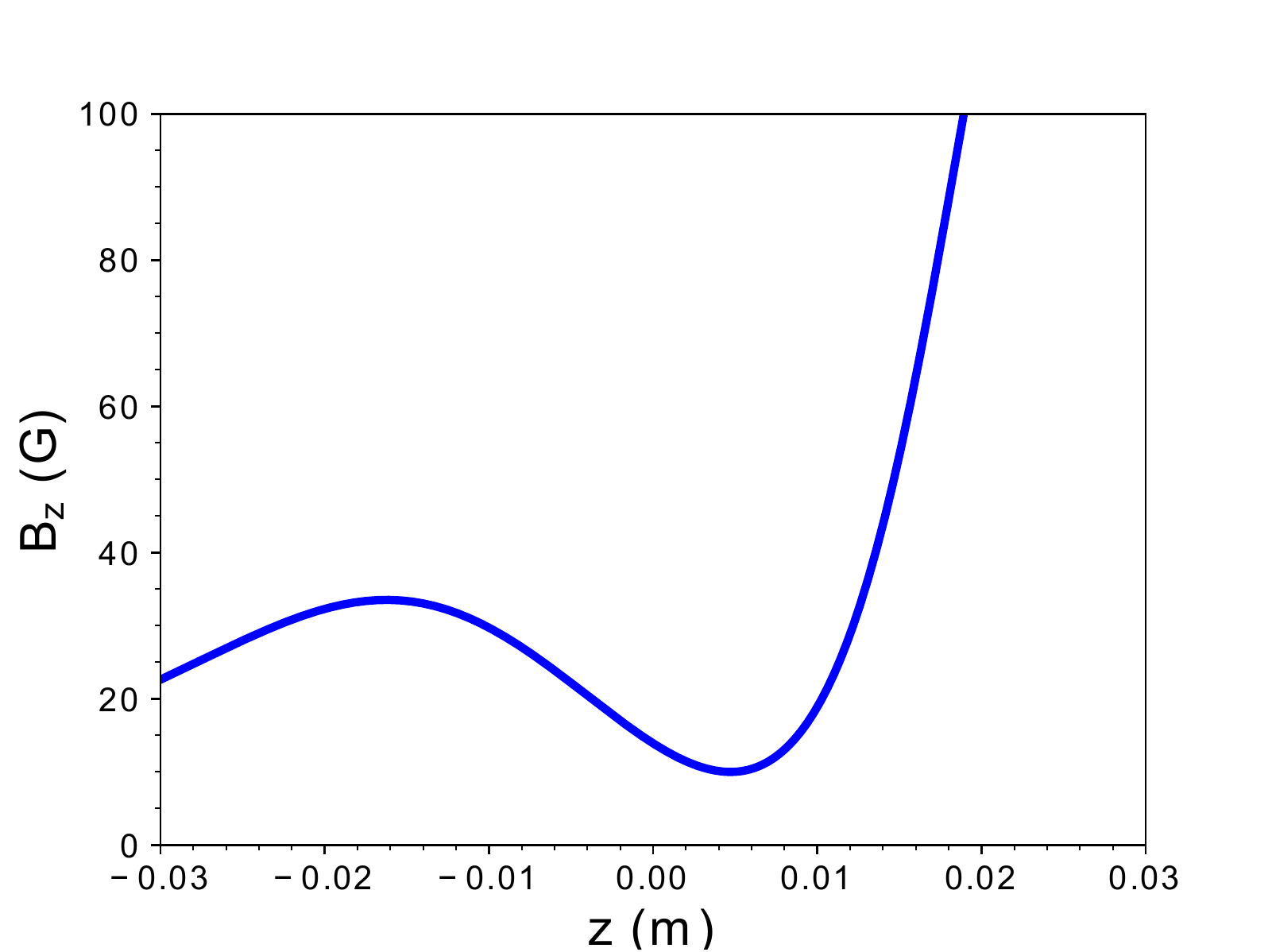}
    \caption{The magnetic trap potential along the vertical axis $\hat{z}$ for our in-vacuum magnetic trap. The trap minimum is 10 Gauss located 6 mm above the centre of the quadrupole coils, while the total trap depth is $\sim$ 25 Gauss.}
    \label{fig:2}
\end{figure}

A number of magnetic trap designs are used in BEC experiments, with some commonly used ones including the cloverleaf design \cite{Mewes96}, the Quadrupole-Ioffe configuration (QUIC) trap \cite{Esslinger98} and atomic chip traps \cite{Fortagh07}.  Any design has strengths and weaknesses, and inevitably involves compromises and trade-offs between factors such as confinement, depth, stability, switch off time, physical size, current required and heat dissipation.  In our magnetic trap setup, no evaporation is implemented; we only perform 1D Doppler cooling in the trap. Hence we require a large enough bias to split the atomic energy levels (as well as prevent Majorana and Penning losses), but are not so concerned about the usual consideration of having a tight confinement, as all evaporation is conducted in the dipole trap. The trap is in a simple QUIC configuration, consisting of two 5 turn anti-Helmhotz (AH) coils of 40 mm diameter centred on the $y$ axis and an 8 turn bias coil of 20 mm diameter centred on the $z$ axis, as shown in Fig. \ref{fig:1}.  These coils are made from 2 mm hollow copper tubing (internal diameter 0.4 mm) mounted in-vacuum to ensure they are close to the atoms.  Cooling is provided by pumping chilled water through the tubes, with each tube attached to hollowed out $1/4$ inch feedthroughs sealed with vacuum compatible solder \cite{solder_note}.  By driving $120$A through the AH coils and $62$A though the bias, we were able to produce a trap of $\sim 1.6$mK depth and $10$G bias field (see Fig. \ref{fig:2}) in the trap centre, which is offset 6mm vertically along the $z$ axis from the centre of the AH coils.  The radial and axial trap frequencies are $\omega_\perp = 2\pi  \times 89$ Hz in the radial axes and $\omega_z = 2\pi  \times 57$ Hz respectively.

A cold source of He$^*$ atoms suitable for loading into a MOT is generated from our He$^*$ beamline, shown schematically in Fig. \ref{fig:1}.  Helium atoms are excited into the metastable state via a hollow cathode DC discharge source \cite{Swansson04}, which is cryogenically cooled using liquid nitrogen. The expanding atomic beam is then collimated using an optical collimation stage featuring 4 beams propagating perpendicular to the He$^*$ atoms with an intensity $\sim 67I_{sat}$ ($I_{sat}  = 0.167 $ mW/cm$^2$) and detuned by $ \sim - 5 \Gamma$ ($\Gamma/2\pi= 1.63$MHz) from resonance.  Note that all detunings for laser cooling and imaging in this work are relative to the $2^3S_1 \rightarrow 2^3P_2$ transition, with a negative sign indicating red detuning.  All laser cooling and imaging beams are sourced from a home built external gain-chip laser \cite{Shin16} with linewidth $<$ 100kHz, which is used to seed a 5W fibre amplifier.  Individual frequency shifts are provided by acousto-optic modulators in each cooling beam. A Zeeman slower is then employed to reduce the velocity of the atoms to $< 100$m/s, featuring a beam of $\sim 92$ $ I_{sat}$ intensity and $\sim -160 \Gamma$ detuning, counter-propagating with the direction of the atomic beam.  A spatially varying field to keep the atoms on resonance while they slow down is generated by two separate coils with variable windings along their length and a zero field in the centre.  These atoms are then trapped and cooled by the 1st MOT, with the 3 counter-propagating MOT beams having $\sim 87$ $I_{sat}$ intensity (horizontal beam), $140$ and $110$ $I_{sat}$ intensity (vertical beams), all with $\sim -22$ $\Gamma$ detuning.  However, the background pressure from the source is too high for this chamber to be used to form a BEC.  The horizontal MOT retro mirror located inside the vacuum chamber has a $\sim 1.5$mm diameter hole in the centre, which enables atoms to leak out of the MOT and into the UHV $2^{nd}$ MOT chamber, where the pressure is $<5\times10^{-11}$Torr.  To assist this process, an additional `push' beam is added ($\sim 110$ $I_{sat}$ intensity and $\sim 3.3 \Gamma$ blue detuned), forming a low velocity intense source scheme (LVIS) \cite{Dall2007}.  The flux from the LVIS, as measured $\sim$10 cm beyond the location of the $2^{nd}$ MOT, is $\sim 9 \times 10^8$ atoms/s, measured using a Faraday cup on a rotation stage and a picoammeter.

The second MOT consists of a quadrupole magnetic field generated by our in-vacuum AH coils with a gradient along the tight axis of $B^{'} \sim 4.4 $G/cm, along with three pairs of counter-propagating laser beams with detunings $\Delta \sim -33 \Gamma$ and intensities of $\sim 140 $ $I_{sat}$,  $\sim 37$ $ I_{sat}$ and $ \sim 400$ $ I_{sat}$. In 1s we load a cloud with $\sim 1.7 \times 10^8 $ atoms at a temperature of $\sim 2.5$mK.  The atom number is measured using saturated fluorescence on an InGaAs photodiode \cite{Dall2007}.  To measure the temperature we switch off the MOT and allow the atoms to fall $\sim 451$mm onto a detector comprising an $80$mm diameter stacked pair of multi-channel plates (MCP).   By using pick-off electronics we extract a signal pulse from the charge depletion of the plates that each atom causes.  After fast amplification and discrimination of the pulses, they are recorded using a digital counter, which provides an integrated 1D time-of-flight (TOF) distribution of the atoms.  The TOF distribution is fitted to a Maxwell-Boltzmann distribution, given by \cite{Yavin02,Dall2007} 
\begin{equation}
n(t) = A v_0^2 \pi \left(\frac{(gt^2/2+l_0)}{t^2}\right) \exp{\left(-\frac{(gt^2/2 - l_0)^2}{v_0^2 t^2}\right)}
\label{TOF_eqn}
\end{equation}
where $A = \left(m/2\pi kT\right)^{3/2}$, $v_0=\sqrt{(2kT/m)}$, $m$ is the mass of a $^4$He atom, $k$ is Boltzmann's constant and $l_0$ is the falling distance. From this distribution we extract the temperature $T$, with the only other free fit parameter being the amplitude $A$. 

Following loading, the MOT is then compressed by ramping the magnetic field down to $1.4$G/cm in $10$ms, while simultaneously the laser detuning is ramped up to $\Delta  \sim - 0.4 \Gamma$ and the intensity of the MOT beams are reduced by two orders of magnitude. The MOT beams are then switched off and the atoms in the $m_J=+1$ state remain trapped in the quadrupole magnetic field of the AH coils. To improve the trap depth and atom density we tighten the quadrupole field to  $16.6$G/cm in $100\mu$s. The bias coil current is then ramped up in $100$ms to form a QUIC trap with a $10$G field offset, removing the magnetic field zero in the quadrupole trap. This configuration provides weak trapping frequencies of $\omega_r \sim 2\pi \times 89$Hz in the radial (x-y) direction and $ \omega_z \sim 2\pi \times 57$Hz in the axial (z) direction. At this stage we have $\sim 5.3 \times 10^7$ atoms at $\sim 0.44$mK, with the lower temperature mostly due to the reduced magnetic trap depth compared to the MOT. The atom number and temperature in the magnetic trap are measured via absorption imaging, with the imaging beam along the x axis being imaged on an InGaAs CCD camera.  Mechanical flipper mirrors allow us to swap between imaging and MOT beams on the same axis, although this prevents absorption imaging during the operation of the MOT.

To cool the cloud further it is illuminated with a 1D Doppler cooling beam \cite{Tychkov06} of intensity $\sim 0.01$ $I_{sat}$, aligned vertically along the bias field axis, circularly polarised to drive the $\sigma^+$ transition and red detuned by $\sim -\Gamma/2$. This Doppler cooling stage cools the cloud in $500$ms, after which we have $\sim 4.9 \times 10^7$  atoms at $\sim 83\mu$K.  This is sufficiently cold to directly transfer the atoms from the magnetic trap to the crossed optical dipole trap.

\begin{figure*}[hbt!]
  \includegraphics[width =1\textwidth]{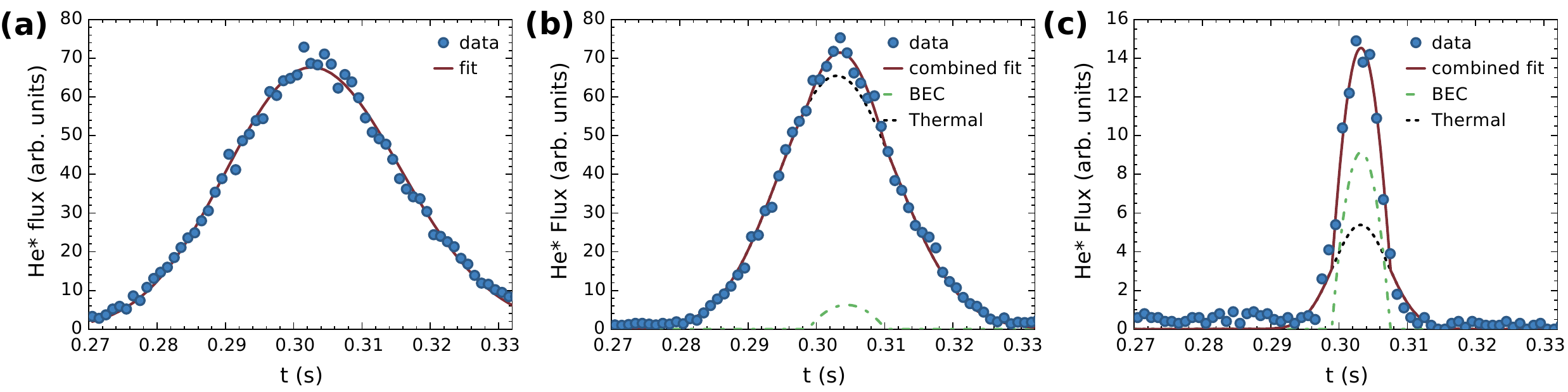}
  \caption{Time of flight signal traces taken from the MCP after the dipole trap is switched off and atoms are allowed to fall onto the detector.  The three plots show different points in the evaporation sequence.  Data is shown as blue circles, with fits shown in red.  Above $T_c$ the fit is to a Maxwell-Boltzmann distribution (see text for details), while below $T_c$ a parabolic Thomas-Fermi fit to the BEC component (green dot-dash line) is added to the thermal fit (black dashed line).  The conditions for the plots shown are: (a) reduced temperature $T/T_c$ = 1.02(5), trap frequencies $\omega_{x,y,z}\sim2\pi\times(1100,920,1400)Hz$, total atom number $N = 3.3(2)\times10^6$.  (b) $T/T_c$ = 0.88(7), $\omega_{x,y,z}\sim2\pi\times(720,590,940)Hz$, $N = 1.7(2)\times10^6$, condensate atom number $N_0=0.2(1)\times10^6$.  (c) $T/T_c$ = 0.54(6), $\omega_{x,y,z}\sim2\pi\times(320,160,360)Hz$, $N = 1.5(5)\times10^6$, $N_0=1.2(4)\times10^6$. Each images represents data from 20 shots
   }
  \label{TOF_plots}
\end{figure*}
\par The final trapping stage is a crossed optical dipole trap (CODT), formed from two intersecting laser beams far detuned from the 1083nm helium resonance at $1550$nm wavelength, from a 100kHz linewidth seed laser that is amplified to $30$W of power.  The power is distributed between the two beams such that after feedback loops to regulate the intensity and losses due to AOM and fibre coupling the first beam (aligned along the $y$ axis) has up to 6W and the second beam (aligned $\sim$ $10 ^\circ$ from the $x$ axis in the $x-y$ plane) has up to 1W of power.  These beams are focused down to Gaussian $1/e^2$ waists of $73\mu$m and $55 \mu$m, respectively, at the location of the atoms.  The two beams intersect at the centre of the $10$G biased QUIC trap, forming a CODT with trapping frequencies $\omega_{x,y,z}\sim2\pi \times (1.1, 1.2, 1.7)$kHz and a trap depth of $\sim 150\mu$K at full power.

Since the 1D Doppler cooling process is more efficient at higher densities \cite{Tychkov06}, we observe that the Doppler cooling is more effective with the dipole beams on, in addition to the magnetic trap.  Here the dipole beams mostly serve to increase the density of the atomic cloud, making the cooling process more efficient and reducing the final temperature after 1D Doppler cooling. In our experiment, this improves the final atom number by $\sim 50 \%$.  Hence the dipole beams are linearly increased to full power over 100 ms prior to the start of the 1D Doppler cooling stage. To transfer the atoms from the QUIC to the dipole trap, after the 1D Doppler cooling stage the trap is linearly decreased in $100$ms to a non-biased trap with $\sim$ 1.1G/cm gradient along the tight axis, before being abruptly turned off with a FET switch.  To preserve the quantisation axis of the atoms and prevent any spin flips during this switch off, an additional large ($\sim$50 cm diameter square) coil mounted above the vacuum chamber was used to generate a uniform magnetic field of $\sim$ 1.6 G aligned along the $\hat{z}$ direction.  This field is switched on before the current through the QUIC trap starts ramping down.  However, once in the dipole trap, it was found that the background magnetic field (dominated by the $z$ component of the Earth's magnetic field) was sufficient to preserve the quantisation, so the extra quantisation coil was switched off 700 ms after transfer to the CODT.  Immediately after transfer we measure $\sim 5 \times 10^6$ atoms at $13.5(1) \mu$K with a phase space density of $\sim 0.5$ in the CODT, an excellent starting point for runaway evaporative cooling. 
\par Evaporation is performed in the CODT by lowering the trap depth through reducing the power of the dipole beams exponentially over 1 second.  As the hotter atoms escape and the remaining atoms rethermalise, this reduces the temperature and enhances the phase-space density.  The evaporation duration is set by the relatively high trap frequencies (and thus high rethermalisation rates).  Unlike in a previous similar He$^*$ experiment  \cite{Bouton15}, we did not observe any improvement in the evaporation process by adding a gradient to cancel the weak components of the dipole potentials along the beam propagation directions. 

To determine the important parameters of the gas in the CODT, we switch off the dipole trap and allow the atoms to fall under gravity onto the MCP to measure the 1D TOF profile along the z axis, integrated in the x-y dimensions.  Examples of the resulting integrated 1D TOF profiles are shown in Fig. \ref{TOF_plots}.  Above the condensation critical temperature $T_c^*$ (note that $T_c$ refers to the non-interacting critical temperature, while $T_c^*$ is the critical temperature corrected for interactions \cite{Dalfovo99}), the profile is fitted to the same Maxwell-Boltzmann distribution as in Eqn. \ref{TOF_eqn} to yield the temperature $T$.  Below $T_c^*$ we fit with a bi-modal distribution consisting of the same Maxwell-Boltzmann distribution plus an inverted parabola to represent the far-field Thomas-Fermi density profile of the BEC \cite{Dalfovo99}.  The half width of this parabola is the far field Thomas-Fermi radius $R_{TF}$, from which we can extract the chemical potential $\mu$ from $\mu\approx m R_{TF}^2/2t_{TOF}^2$ \cite{TOF_note}.  From the chemical potential we can extract the number of atoms in the condensate $N_0=(2\mu)^{5/2}/15\sqrt{m}\hbar^2\bar{\omega}^3a$, where $a$ is the He$^*$ $s-$wave scattering length.  The number of atoms in the thermal cloud $N_T$ is given by $N_T\approx1.202\left(k_B T/\hbar \bar{\omega}\right)^3$, for $T<T_c^*$. 
To extract the atom number above $T_c^*$, we sum the integrated counts, correct for the fraction of the cloud at temperature $T$ that will hit the detector and then scale by an effective quantum efficiency of the detector $\gamma_{QE}$.  We vary $\gamma_{QE}$ until the thermal atom number $N_T$ matches for the clouds just above and below $T_c^*$.  Note that non-linearities in $\gamma_{QE}$ at high fluxes mean this is not accurate for $N_0$.
 
\begin{figure}[hbt!]
    \centering
    \includegraphics[width=\linewidth]{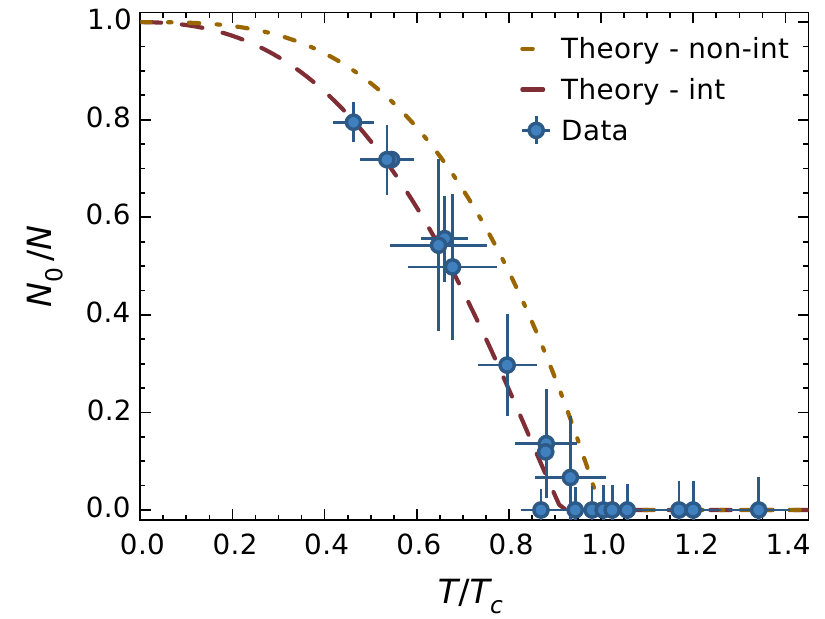}
    \caption{Condensate fraction $N_0/N$ verses reduced temperature $T/T_c$ for different points in the evaporation sequence.  Data is shown as blue points, with error bars dominated by fit uncertainties due to the condensate and thermal widths being relatively similar for many data points, making accurate fits difficult.  The brown dot-dash line shows the non-interacting theory, while the red dashed line shows a correction accounting for two-body interactions \cite{Dalfovo99}.  For $N_0/N \gtrsim 0.8$ the thermal component is obscured by the condensate, which prevents reliable fits, hence the final evaporation points are not shown.}
    \label{N0_Tc}
\end{figure}

The above information, along with the total atom number $N=N_0 + N_T$ and $k_BT_c=0.94\hbar\bar{\omega}N^{1/3}$, is then combined to produce Fig. \ref{N0_Tc}.  This shows the condensate fraction $N_0/N$ vs the reduced temperature $T/T_c$ for our sequence as we evaporate through the transition temperature.  We cross $T_c^*$ at $\sim 6\mu K$ with $\sim 3 \times 10^6$ atoms, and by evaporating further we are able to produce BECs with no discernible thermal fraction (not shown on the graph) with $\sim 1 \times 10^6$ atoms.  

We demonstrate a comparatively simple and short method for producing Bose Einstein condensates of He$^*$ atoms. By using a magnetic trap with a 10G bias generated from an in vacuum set of coils, we are able to cool the cloud below $100$  $\mu K$ in the magnetic trap via 1D Doppler cooling. The degenerate state of metastable helium atoms is accomplished via direct evaporative cooling in a crossed optical dipole trap.  We cross the condensation critical temperature with $\sim3 \times 10^6$ atoms at $\sim 6\mu$K and by evaporating further are able to generate essentially pure BECs with  $\sim 1 \times 10^6$ atoms. The entire sequence to produce a BEC only takes 3.3 sec.  This will provide an excellent starting point for a range of experiments with He$^*$ BECs, such as many-body correlation experiments \cite{Hodgman2017}, probes of quantum non-locality \cite{Shin2018} or strongly interacting optical lattice physics \cite{Cayla20}.

\begin{acknowledgments}
 The authors would like to thank David Clement and Robert Dall for helpful discussions, and Colin Dedman, Ross Tranter and Liangwei Wang for technical assistance. This work was supported through Australian Research Council (ARC) Discovery Project grants DP160102337 and DP190103021. SSH was supported by ARC Discovery Early Career Researcher Award DE150100315.
\end{acknowledgments}
  
\bibliography{Refs}

\begin{thebibliography}{39}
\bibitem [{sol()}]{solder_note}%
  \BibitemOpen
  \href@noop {} {}\bibinfo {note} {Note that standard solder and flux was also
  found to work, although required an initial period under vacuum to remove the
  residual flux before being installed in the final UHV chamber. This prevented
  out-gassing from the non-vacuum compatible flux, was sufficient to still be
  able to reach UHV pressures.}\BibitemShut {Stop}%
\bibitem [{TOF()}]{TOF_note}%
  \BibitemOpen
  \href@noop {} {}\bibinfo {note} {While this approximation for the chemical
  potential from the far field $R_{TF}$ is only valid for condensates in cigar
  shaped traps, numerical solutions of the equations of motion following
  \cite{Castin96} show that it gives at most an $\approx 10\%$ error in $\mu$
  for the trap geometries in our system, which is within our experimental
  uncertainties.}\BibitemShut {Stop}%
\end{thebibliography}

\end{document}